\documentclass[aps, twocolumn, superscriptaddress, amsmath, amssymb, floatfix, prb, longbibliography]{revtex4-2}

\usepackage{amsfonts}
\usepackage{graphicx}
\usepackage{enumerate}
\usepackage{bm}
\usepackage{bbm}
\usepackage{color}
\definecolor{prlblue}{rgb}{0.18,0.19,0.57}
\usepackage{braket}
\usepackage{tikz}
\usetikzlibrary{patterns.meta} 
\usetikzlibrary{calc}
\usepackage[dvipsnames]{xcolor}

\newcommand{\be}{\begin{equation}}
\newcommand{\ee}{\end{equation}}
\usepackage{hyperref}
\usepackage{natbib}
\begin{document}

\title{Entropic Barriers and the Kinetic Suppression of Topological Defects}

\author{Yi-Lin Tsao}
\author{Zhu-Xi Luo}
\affiliation{School of Physics, Georgia Institute of Technology, Atlanta, Georgia 30332, USA}

\date{\today}

\begin{abstract}
Many quantum phases, from topological orders to superfluids, are destabilized at finite temperature by the proliferation and motion of topological defects such as anyons or vortices. Conventional protection mechanisms rely on energetic gaps and fail once thermal fluctuations exceed the gap scale. Here we examine a complementary mechanism of entropic protection, in which defect nucleation is suppressed by coupling to mesoscopic auxiliary reservoirs of dimension 
$M$, generating an effective free-energy barrier that increases with temperature. In the Ising chain, this produces a characteristic three-regime evolution of the correlation length as a function of temperature — linear growth, entropy-controlled plateau, and eventual breakdown—indicating a general modification of defect behavior. 
Focusing on two spatial dimensions, where true finite-temperature topological order is forbidden in the thermodynamic limit, we show that entropic protection can nevertheless strongly enhance stabilization at finite system size, the regime directly relevant for quantum memory and experiments. 
Owing to the topological character of the defects, creation and transport are independently suppressed, yielding a double parametric reduction of logical errors in the entropic toric code and enhanced coherence when the framework is extended to Berezinskii–Kosterlitz–Thouless transitions. Entropic barriers thus provide a passive and scalable route to stabilizing quantum phases in experimentally relevant regimes. We propose an experimental setup for entropic toric code using dual-species Rydberg arrays with dressing.
\end{abstract}

\maketitle

\setcounter{tocdepth}{1} 
{
  \hypersetup{linkcolor=magenta}
  \tableofcontents
}

\section{Introduction}

Understanding how quantum phases persist or degrade at finite temperature is a central challenge in condensed matter physics and quantum information science. In two spatial dimensions, while topological order offers intrinsic protection against local perturbations \cite{KITAEV20032}, this protection is generically compromised by the proliferation of topological defects \cite{PhysRevB.76.184442, PhysRevB.77.064302}. In discrete gauge theories like the toric code, thermally activated point-like anyons diffuse and dephase logical memory \cite{PhysRevB.76.184442, PhysRevB.77.064302,RevModPhys.88.045005}. Similarly, in more conventional settings of two-dimensional superfluids, coherence is destroyed by the unbinding of vortex-antivortex pairs (BKT transition) \cite{kosterlitz1973ordering}. The instability in both cases stems from the reliance on purely energetic protection: once the system’s extensive configurational entropy overcomes the static energetic tension of these defects, they proliferate and destroy global order.

Mitigating this fragility traditionally necessitates a trade-off between the substantial redundancy requirements of active quantum error correction \cite{PhysRevA.52.R2493,Dennis_2002,gottesman1997stabilizer,RevModPhys.87.307} and the geometric intractability of energetic self-correction via high-dimensional or fractal confinement  \cite{Dennis_2002,doi:10.1142/S1230161210000023,PhysRevA.83.042330,PhysRevLett.107.150504}. A promising alternative is autonomous quantum error correction, in which quantum information is stabilized through engineered dissipation without measurement-feedback loops \cite{PhysRevA.91.042322,PhysRevX.4.041039}. These protocols suppress errors by continuously removing entropy from the logical subspace \cite{PhysRevA.72.012306}, with the resulting protection being inherently dynamical and logical stability set by non-equilibrium rate balances rather than a thermodynamic free-energy barrier. 
In contrast, our approach fundamentally diverges by exploiting thermal entropy itself as a stabilizing resource, rather than continuously extracting it.

In this work, we examine a universal mechanism of entropic protection, where fluctuating auxiliary degrees of freedom generate entropic forces that stabilize a target order at finite system size through a modified free-energy landscape. Specifically, we couple topological defects to local entropic reservoirs—mesoscopic baths of auxiliary states—engineered such that the defect-free vacuum allows the reservoir to explore a large phase-space volume $M$, while the presence of a defect confines it to a unique state. This asymmetry produces an entropic free-energy barrier which scales intrinsically with temperature and therefore remains effective even when energetic gaps are overwhelmed by thermal fluctuations. 

We apply this framework to both discrete and continuous topological defects. In section \ref{sec:Ising}, to illustrate the idea, we exactly solve the entropic Ising chain and find the behavior of the correlation length to exhibit three distinct regimes as a function of temperature: linear increase, plateau and eventual breakdown. Moving on to the two-dimensional entropic toric code in section \ref{sec:ToricCode}, we analyze both the static and dynamical stabilities of the toric code ground state at finite system size and uncover a ``double suppression'' effect: the entropic bath not only
suppresses the rate of anyon pair creation by $\sim M^{-2}$, but also kinetically suppresses the diffusion of anyons, reducing their mobility by a factor of $M^{-1}$. A concrete implementation using Rydberg atom arrays generalizing the dual-species ``Weimer-bath'' architecture \cite{weimer2010rydberg} is proposed in section \ref{subsec:experiment}. Extending the discussion to the BKT transition, we show in section \ref{sec:XY_Model} that the entropic coupling effectively renormalizes the vortex core energy, restoring a nonzero tension to vortex worldlines and thereby pushing the crossover temperature to significantly higher values at finite system size. 

Our work is strongly inspired by recent studies of entropic stabilization in equilibrium lattice systems~\cite{han2025entropicorder, 2512.07980},
which demonstrated that coupling a system to auxiliary degrees of freedom with a large entropic capacity can evade no-go theorems forbidding long-range order or entanglement at high temperature~\cite{PhysRevX.4.031019,bakshi2025hightemperaturegibbsstatesunentangled}. The references primarily focused on the existence of stable phases in the thermodynamic limit, emphasizing higher spatial dimensions where true finite-temperature order can persist. The present work explores a complementary and experimentally motivated regime. Rather than seeking stabilization in the two-dimensional thermodynamic limit — where topological order remains forbidden—we focus on finite-size systems, which are the natural setting for quantum memories and programmable quantum simulators. We concentrate on the kinetic and dynamical aspects of entropic protection, including defect nucleation, transport, and lifetime scaling, explicitly analyze memory stability in two spatial dimensions, and propose concrete experimental implementations. In this sense, our work provides a dynamical  realization of entropic stabilization tailored to practical quantum information processing. Finally, our mechanism is distinct from the phenomenon of order by disorder~\cite{villain1980order}. In those scenarios, thermal fluctuations select a specific ordered state from a degenerate manifold by maximizing entropy within the low-energy sector. In contrast, entropic protection relies on neither intrinsic frustration nor ground-state degeneracy. Instead, entropy is engineered as an auxiliary barrier that suppresses the nucleation and motion of defects, hardening the phase against thermal erosion. While order by disorder uses entropy to select an ordered state, our mechanism uses entropy to protect it.

While not directly addressing our setup, related discussions of protected order at finite or high temperatures can also be found in Refs.~\cite{10.1063/1.1794652,PhysRevE.63.031503,GOLDSTEIN1959390,YOSHIDA20112566,PhysRevLett.125.131603,PhysRevD.102.065014,10.21468/SciPostPhys.14.6.168,Chai_2021,Chaudhuri_2021,PhysRevD.103.096014,Chaudhuri_2021, Agrawal_2021, PhysRevLett.134.041602, yb7d-6tvc, PhysRevD.103.026008, BUCHEL2021115425, BUCHEL2021115605, BUCHEL2024116578, 10.21468/SciPostPhys.12.6.181, Meng,li2019finitetemperaturetopologicalentanglemententropy,WEINSTEIN2020168018,PhysRevResearch.4.L032008}.

\section{The Entropic Ising Chain}
\label{sec:Ising}

This warm-up section considers the Ising model coupled to local entropic reservoirs with dimension $M$.  In standard ferromagnetic Ising chain, the correlation length is purely energetic, $\xi \sim e^{2\beta J}$, and diverges only at zero temperature ($\beta\rightarrow \infty$). 
Below we will show that entropic reservoirs generate a correlation length $\xi \propto M$ that, when $M$ is comparable with system size $L$, can remain macroscopic even at high temperatures.  
The Hamiltonian is:
\begin{equation}
    H = \sum_{i} \hat{n}_i  \left( \epsilon \frac{1+Z_i Z_{i+1}}{2} + J \frac{1-Z_i Z_{i+1}}{2} \right).
\label{eq:Ising}
\end{equation}
Here, $\hat{n}_i$ describes a local bath with $M$ available states ($n_i=0,1,\dots,M-1$). The interaction enforces a state-dependent phase space volume: If a domain wall exists ($Z_i Z_{i+1} = -1$), the bath experiences a potential via $J \hat{n}_i$. When the potential is steep $\beta J \gg 1$, the bath is confined to its ground state where $\langle \hat{n}_i\rangle =0.$
Conversely, if the stabilizer is satisfied ($Z_i Z_{i+1} = +1$), the small $\epsilon$ allows the boson to effectively delocalize over a large phase space volume in the ground sector of spins while regulating the bath states to be normalizable. 

\begin{figure}[htbp]
\centering
\includegraphics[scale=0.49]{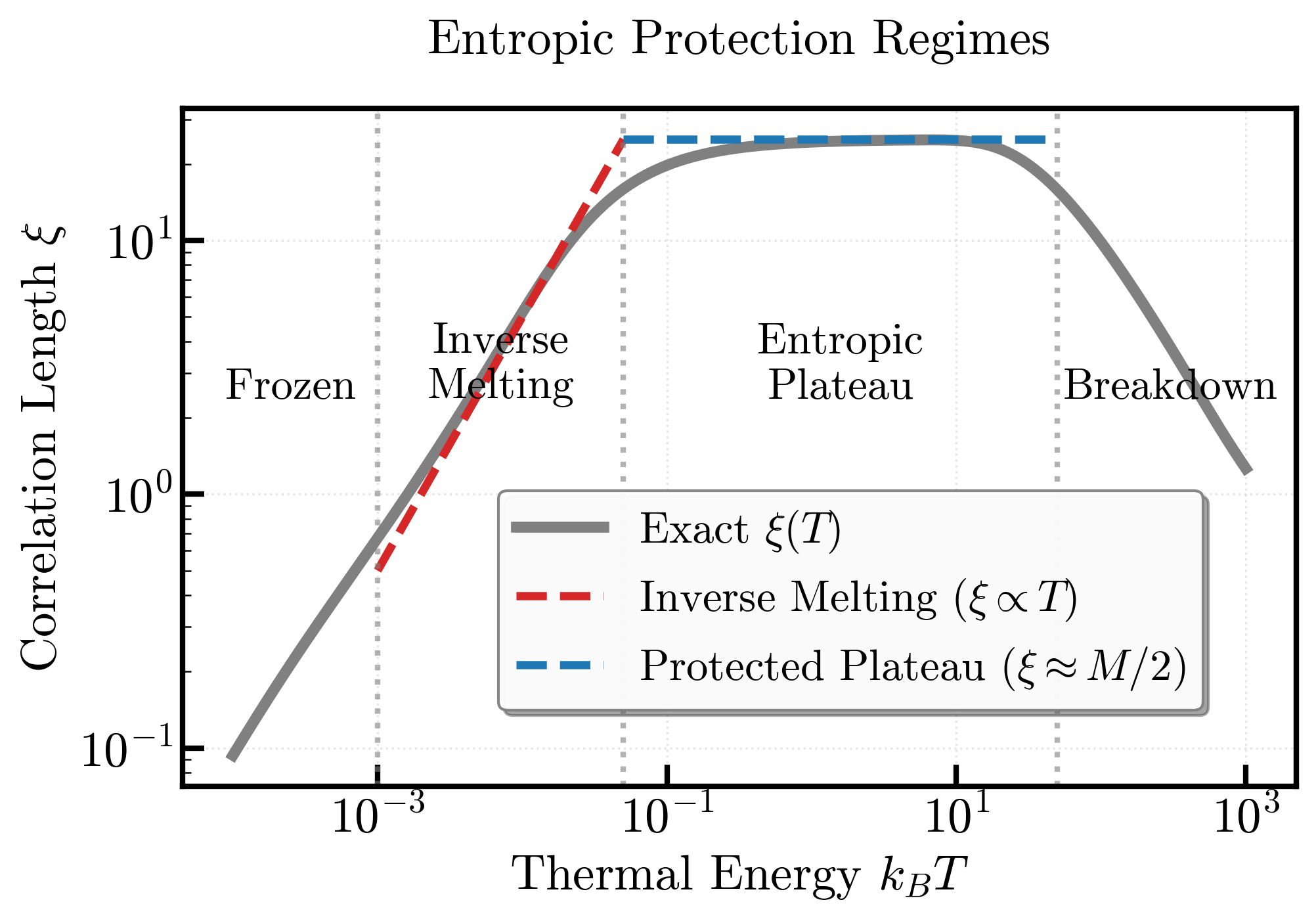}
\caption{Correlation length as a function of thermal energy $1/\beta$. Parameters are chosen as $M=50$, $\epsilon=10^{-3}$, $J=50$. }
\label{fig:Ising}
\end{figure}

The spin correlation length can be exactly solved by integrating out the bath. The full partition function $Z = \text{Tr}(\prod T_i)$ is determined by the transfer matrix in the computational spin basis:
$T = \begin{pmatrix} w_+ & w_- \\ w_- & w_+ \end{pmatrix},$ where $w_{\pm}$ are the Boltzmann weights for the local link configurations $Z_i Z_{i+1}=\pm 1.$ Since the eigenvalues of $T$ are $w_+\pm w_-,$ it's straightforward to derive  the spin-spin correlation function, which decays as:
\begin{equation}
    \langle Z_i Z_{i+r} \rangle = \left( \frac{w_+ - w_-}{w_+ + w_-} \right)^r.
\end{equation}
Upon identification of the correlator with the form $e^{-r/\xi}$, one can derive the correlation length to be 
\be
\xi =\left[ \ln \frac{\lambda +1}{\lambda -1}\right]^{-1}, 
\ee
where $\lambda$ is the ratio 
\begin{equation}
    \lambda \equiv \frac{w_+}{w_-} = \frac{1-e^{-\beta \epsilon M}}{1-e^{-\beta \epsilon}} \frac{1-e^{-\beta J}}{1-e^{-\beta J M}}.
\label{eq:lambda}
\end{equation}
There are three regimes with distinct physics as $\beta$ decreases, when the thermal energy is larger than $\epsilon$. The results are summarized in fig. \ref{fig:Ising}. 
\textit{Inverse melting regime}: $ \beta \epsilon \ll 1 \ll  \beta \epsilon M$; $\beta J\gg 1$. The thermal energy exceeds the boson level spacing $\epsilon$, allowing the bath to behave as a thermal gas in the ground sector of spins. However, the temperature is still lower than the total energy span $M\epsilon$, so the bath does not yet feel the hard cutoff $M$. In this case, $w_+ \approx (\beta \epsilon)^{-1}$, while $w_- \approx 1$ is still frozen by $J$. 
The correlation length thus scales as  $\xi\sim (\beta \epsilon)^{-1}$, which grows linearly with temperature.

\textit{Saturated entropic plateau}: $\beta \epsilon M\ll 1 \ll \beta J$. 
The temperature is now high enough to saturate the entire bandwidth of the bath, so all $M$ states are equally accessible. Yet, it is still too low to overcome the defect penalty $J$. Now $w_+ \approx \beta \epsilon M/\beta \epsilon = M$ and $w_- \approx 1$ is still frozen. In this case, $\xi \approx M/2$. The correlation length hits a robust maximum plateau and becomes temperature independent. In the thermodynamic limit $L\rightarrow \infty$, for any fixed finite $M$, true long range order is absent. Nevertheless, at finite $L$, the system behaves as if it were Ising ordered once $M$ is comparable to $L.$

\textit{Breakdown regime}: $\beta J\ll 1$. When the thermal fluctuations finally exceed the defect energy $J$, $w_+$ remains saturated at $M$, but $w_-$ begins to grow as $1/\beta J$ when the defect bath gets excited. 
Thus
$\lambda\simeq MJ/k_BT$ and $\xi$ falls as $1/T$, so a chain of length $L$
retains $\xi\ge L$ up to
$
k_BT^{*}(L)\simeq{MJ}/{2L},
$
which exceeds the plateau edge only if $M>2L$. 

We remark that at strictly zero temperature, $\beta\to\infty$, the bath is frozen into its ground state $n_i=0$ for all spin configurations, and the Hamiltonian~\eqref{eq:Ising} vanishes identically. As a result, all spin configurations are exactly degenerate and the spins are completely disordered. The ordering mechanism discussed here is therefore intrinsically a finite-temperature effect, arising purely from entropic contributions to the free energy. To connect this entropically stabilized Ising physics to the conventional zero-temperature limit, one can add a direct nearest-neighbor Ising interaction, which will be discussed in Appendix \ref{app:modified_Ising}.

\section{Entropic Protection of 2D Toric Code}
\label{sec:ToricCode}

This section discusses the entropic protection of topological quantum memory in two dimensions. We will focus on the Toric Code \cite{KITAEV20032} defined on a square lattice with periodic boundary conditions. The logical information is encoded in the degenerate ground space absent of anyonic excitations.

The system consists of gauge qubits located on the links of the lattice, coupled to bath degrees of freedom located on the vertices ($v$) and plaquettes ($p$). The Hamiltonian is given by:
\begin{equation}
\begin{split}
H = &\ \frac{1}{2}  \sum_v \left[ J(1-\hat{A}_v) \hat{n}_v +  \epsilon (1+\hat{A}_v) \hat{n}_v \right]\\
& + \frac{1}{2}  \sum_p \left[ J (1-\hat{B}_p) \hat{n}_p + \epsilon (1+\hat{B}_p)  \hat{n}_p \right].
\end{split}
\label{eq:TC}
\end{equation}
The standard toric code stabilizers are $\hat{A}_v = \prod_{\partial j \ni v} X_j$ and $\hat{B}_p = \prod_{j \in p} Z_j$. The operators $\hat{n}_{v,p}$ represent the coupling to local entropic reservoirs.

While an inverse melting regime is also present similar to the Ising case in Sec. \ref{sec:Ising}, we will focus on the saturated entropic regime, $\beta \epsilon M  \ll 1 \ll \beta J$, where the correlation physics is independent of temperature. 
In this window, the partition function for a local reservoir is $w_+ \approx M$ if the stabilizer is satisfied, and $w_- \approx 1$ if it is violated. This generates an effective free energy cost for anyons: the probability ratio of a state with a single anyon pair to the ground state is: $(w_-/w_+)^2 \approx M^{-2}$, which corresponds to an entropic free energy cost of $\beta^{-1} \ln M$ per violated stabilizer. Unlike energetic gaps which are fixed constants, this chemical potential scales linearly with temperature, maintaining a constant suppression of anyon density even as $T$ increases.

\subsection{Static stability}

To quantify the preservation of topological order in the Gibbs state $\rho(T) = e^{-\beta H}/\mathcal{Z}$, this section computes the scaling of Wilson loops as well as the topological entanglement entropy. 

The Wilson loop is defined as usual 
$
W_z (R) = \prod_{p \in R} B_p
$
surrounding a region $R$. 
In the regime of saturated entropic plateau, the expectation value for a single $B_p$ is 
$
\langle B_p \rangle = (w_+-w_-)/(w_++w_-) \sim 1-2/M
$
when expanded in large $M$. The scaling of the Wilson loop is thus 
\be
\langle W_z (R) \rangle =\prod_{p\in R}\langle B_p\rangle \sim e^{|R|\ln (1-2/M)}.
\ee
Note that the logarithm is always negative. Therefore, the Wilson loop is area-law decaying except when $M\rightarrow \infty$, which strictly speaking indicates confinement at any finite $M$. However, we emphasize that neither $\beta$ nor system size $N$ enters the expression. While strictly this implies a confined phase, the entropic confinement is nontrivial and fundamentally different from standard thermal destabilization, as the effective string tension is temperature-independent and can be arbitrarily suppressed by increasing the bath dimension $M$ rather than cooling the system. 

Next we examine the topological entanglement entropy \cite{PhysRevLett.96.110404,PhysRevLett.96.110405} which by construction cancels out classical correlations and can serve as a diagnosis for topological order at finite temperature. 
For each toric code stabilizer $S$ of type either $\hat{A}_v$ or $\hat{B}_p,$ the local density matrix after tracing out the corresponding bosons $\hat{n}_S$ reads, 
\be
\rho_S \propto \frac{1+S}{2} \frac{1-e^{-\beta \epsilon M}}{1-e^{-\beta  \epsilon}} + \frac{1-S}{2} \frac{1-e^{-\beta J M}}{1-e^{-\beta J}}.
\ee
Such a density matrix can be identified with the thermal density matrix of standard toric code $\rho_S =C e^{\beta_{\text{eff}} S}$, where
\be
\beta_{\text{eff}}= \frac{1}{2}\ln \frac{(1-e^{-\beta \epsilon M})(1-e^{-\beta J})}{(1-e^{-\beta \epsilon })(1-e^{-\beta J M})}.
\ee
This establishes a direct mapping to the finite-temperature toric code, whose topological entanglement entropy vanishes at any non-zero temperature \cite{PhysRevB.76.184442} in the thermodynamic limit. For a finite system with $N$ sites, the topological correction to the entropy remains robust only when $\sqrt{N} e^{-\beta_{\text{eff}}} \ll 1$. In our saturated entropic plateau regime, where $\beta_{\text{eff}}\sim \frac{1}{2}\ln M$, this  implies a threshold condition for the bath size of $M\sim N$. Notably, $M$ is the Hilbert space dimension of the bath - if the bath contains $m$ qubits, the number of qubits required for each bath only needs to scale with the logarithm of the number of system qubits 
\be
m\sim \log_2 N. 
\ee
While this scaling implies that $m$ must diverge in the thermodynamic limit to preserve equilibrium order, this is not a practical bottleneck for quantum information processing. For a realistic fault-tolerant memory patch with system size $N=250$, the required bath size $m\approx 8$ is within the reach of current experimental platforms: For example, ref. \cite{Manetsch_2025} demonstrates Cs arrays holding $6100$ atoms in
$\simeq12{,}000$ tweezer sites; species-selective placement of Rb and Cs atoms on generic geometries has also been demonstrated \cite{PhysRevX.12.011040,Anand2024}. 
We will describe a possible experimental proposal in section \ref{subsec:experiment}.

\subsection{Dynamic Stability}

The lifetime of the quantum memory depends on the dynamical coupling to external environment, which we will examine using a Lindblad master equation: 
\begin{equation}
\begin{split}
\dot{\rho} = \Gamma_0 \sum_l \bigg( & \frac{1}{M^2} \mathcal{D}[L_{\text{cr}, X_l}] + \frac{1}{M} \left( \mathcal{D}[L_{\text{dif}, X_l}^{\to 1}] + \mathcal{D}[L_{\text{dif}, X_l}^{\to 2}] \right) \\
& + \mathcal{D}[L_{\text{ann}, X_l}] + (X_l \to Z_l) \bigg) \rho.
\end{split}
\label{eq:Lindblad}
\end{equation}
Here $\mathcal{D}[L]\rho = L\rho L^\dagger - \frac{1}{2}\{L^\dagger L, \rho\}$, $\Gamma_0$ is the bare rate of each Pauli channel, and the jump operators $L$'s can be expressed by local Pauli operator $X_l$ and $Z_l$'s, as well as $A_v$, $B_p$ projectors, see Appendix \ref{app:Lindblad} for detailed derivation of \eqref{eq:Lindblad}. This master equation reveals a distinct hierarchy of transition rates:

\textit{Pair Creation.} The term $\mathcal{D}[L_{\text{cr}}]$ describes the nucleation of a defect pair from the vacuum. This process requires two local reservoirs to spontaneously fluctuate from the flat sector (volume $M$) into the confined sector (volume 1), incurring a rate penalty $\Gamma_{\text{cr}} \sim M^{-2}$. 

\textit{Anyon Diffusion.} The terms $\mathcal{D}[L_{\text{dif}}]$ describe the hopping of an anyon to an adjacent site. While energetically neutral, this transport is kinetically gated. The reservoir at the target site must fluctuate into a specific microstate to accept the defect, reducing the effective diffusion constant to $D_{\text{eff}} \sim D_0 M^{-1}$.

\textit{Pair Annihilation.} The process $\mathcal{D}[L_{\text{ann}}]$ relaxes the baths into the high-entropy vacuum sector and is therefore entropically uninhibited, proceeding at the bare rate $\Gamma_0$.

A key question for passive quantum memory is how the lifetime of the logical information depends on the linear system size $L= \sqrt{N}$  and on the entropic reservoir dimension $M$. A logical error occurs when the pair of anyons created winds around a non-contractible loop of the torus before getting annihilated again. The rates in \eqref{eq:Lindblad} satisfy detailed balance with respect to $e^{-\beta H}$, so that the ratio between probabilities for two neighboring excitations satisfies 
$P(-,-)/P(+,+)=\Gamma_{\rm cr}/\Gamma_{\rm ann}=M^{-2}$, which is a two-site quantity. The excitation density can then be solved as $n/(1-n) = M^{-1}$ where $n$ is the anyon density, and the average separation between excitations is
$\ell=n^{-1/2}\simeq\sqrt M$. 
The crossover to a finite anyon density lies at $L\sim\ell\simeq \sqrt{M}$. Since static
stability requires $M\gtrsim L^{2}$, the system never reaches a finite-density regime while topological order survives;  we therefore focus on the defect-rare regime.

A logical error requires a nucleated pair to wind before annihilating. Pairs of either type of excitations nucleate on any of the $2L^{2}$ links at rate $\Gamma_{\rm cr}$. A
nucleated pair initially sits at contact. Next, while pair annihilation is entropically unobstructed
at $\Gamma_{\rm ann}$, pair separation is gated: each defect can hop 
across three links other than the shared one at $\Gamma_{\rm dif}$. Therefore,  the probability of separating rather than annihilating is $6\Gamma_{\rm dif}/(6\Gamma_{\rm dif}+\Gamma_{\rm ann})\simeq6/M$.  
Once separated, the relative coordinate between the pair goes through a random walk and reaches separation $\sim L$ before returning to contact with probability $\ln2/\ln(L/2)$. As for pairs
that return, they restart the process, which alters the result only at higher order in $1/M$. Combining this nucleation-escape-winding process, the lifetime can be derived as
\begin{equation}
\tau_L\simeq\frac{M^{3}\ln(L/2)}{c\,\Gamma_0L^{2}},\qquad c\approx8.
\label{eq:TC_lifetime}
\end{equation}
Since $X$- and $Z$-errors are decoupled, logical errors occur at the total rate $2/\tau_L$ when both channels are active, i.e.\ $c\to2c$, with the scaling unchanged. 
The three powers of $M$ are the double suppression: $M^{2}$ from the entropic
cost of nucleating a pair, and one further power from the kinetic gating of
transport. At fixed $M$, the lifetime falls as $L^{-2}\ln L$. Holding $\tau_L$ fixed requires $M\propto(L^{2}/\ln L)^{1/3}$, i.e.\
$m\simeq\tfrac23\log_2L$.

It is instructive to compare this result with the standard toric code, where protection arises purely from an energy gap $\Delta$. In that case, the memory lifetime follows the Arrhenius law $\tau_{\text{std}} \sim \tau_0 \exp(\beta \Delta)$, which is extremely sensitive to thermal fluctuations.

\subsection{Experimental setup }
\label{subsec:experiment}

States in the toric code phase have been prepared and probed on single-species Rydberg arrays~\cite{Bluvstein_2023,doi:10.1126/science.abi8794}, while species-selective placement and control in dual-species arrays have been achieved in \cite{PhysRevX.12.011040, Anand2024}.  Combining the two, we propose a concrete implementation of the entropic toric code using a dual-species Rydberg atom array (e.g., $^{87}\text{Rb}$ for system qubits living on the links and $^{133}\text{Cs}$ for bath atoms living on the vertices and plaquette centers) 
\cite{PhysRevA.92.042710}. 
A central challenge is implementing the conditional Hamiltonian term which applies a steep penalty to the bath only when the 4-body stabilizer is violated. To bridge the locality gap between the system stabilizer and the bath potential, a single auxiliary sensor atom (Cs) needs to be introduced at the center of each plaquette and vertex (in addition to the bath atoms), serving as a mediator. The protocol operates in a hybrid digital-analog regime:

\textit{(i) Stabilizer Mapping.} We utilize the standard ancilla-based stabilizer-measurement circuit, in which the parity of four data qubits is mapped onto a sensor atom via four ancilla–data entangling operations (equivalent to four CNOT or CZ gates at the logical level \cite{gottesman1997stabilizer}).
In Rydberg-atom platforms, such coupling is implemented using native blockade-based pulse sequences rather than explicit two-qubit gates, following the general schemes introduced in \cite{weimer2010rydberg,MULLER20121} and demonstrated experimentally in \cite{PhysRevLett.123.170503,Madjarov_2020}. If the stabilizer is satisfied ($+1$), the sensor will remain in the ground state $|g\rangle$. If violated ($-1$), the sensor will flip to a metastable clock state $|e\rangle$.

\begin{figure}[htbp]
\centering
\includegraphics[width=\columnwidth]{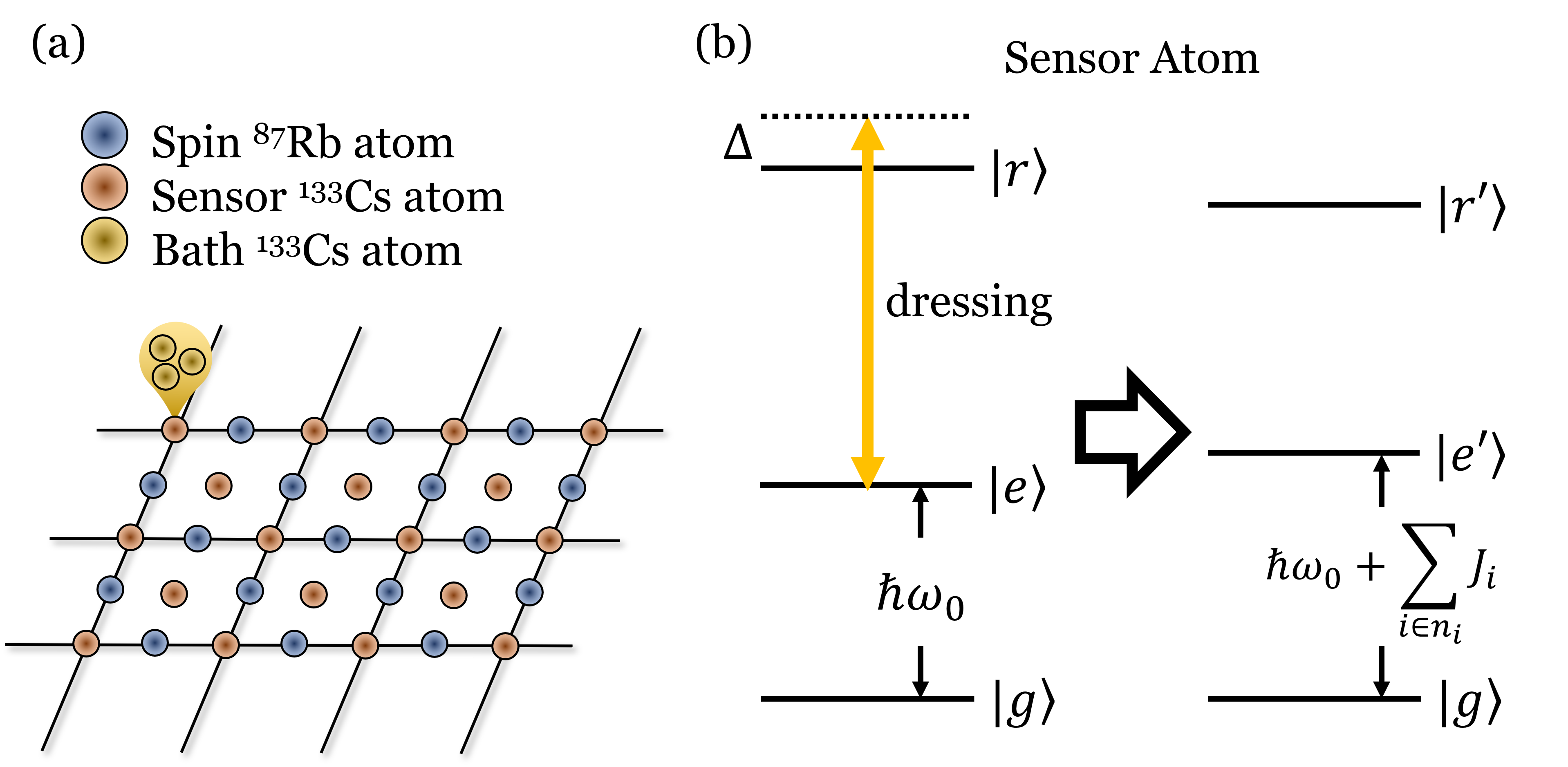}
\caption{(a) Schematic of the proposed dual-species Rydberg atom array, which consists of target spins (blue, $^{87}$Rb), sensor atoms (red, $^{133}$Cs), and bath atoms (yellow, $^{133}$Cs). Each sensor serves as a messenger between the system lattice and a local entropic reservoir formed by the bath cluster. (b) Conditional blockade via Rydberg dressing. The relevant energy levels $|g\rangle, |e\rangle,$ and $|r\rangle$ correspond to the ground, excited, and Rydberg states of the Cs atoms. An off-resonant laser with Rabi frequency $\Omega$ and negative detuning $|\Delta| \gg \Omega$ dresses the excited state, creating $|e'\rangle$ with a small admixture of Rydberg $|r\rangle$. This induces strong van der Waals interactions $\propto J_i$ between the sensor and surrounding bath atom $i$. The resulting total energy shift, $\sum J_i$, enforces the energetic penalty for defects. 
Crucially, the entropic stabilization relies on the degenerate bath configurations rather than precise energy matching, making the protocol robust against variations in interaction strengths $J_i$ arising from different inter-atomic distances $R_i$ between the bath and
sensor atoms.
}
\label{fig:expsetup}
\end{figure}

\textit{(ii) Conditional Blockade.} To generate the steep defect penalty $J$, we use the technique of Rydberg dressing \cite{PhysRevA.82.033412,PhysRevLett.104.195302}. We propose to apply an off-resonant laser field that couples the sensor's excited state $|e\rangle$ (and consequently the bath atoms) to a high-lying Rydberg state $|r\rangle$. This laser does not fully excite the atoms to the Rydberg state (which would decay too quickly); instead, it dresses the ground state by mixing in a small fraction of Rydberg character. This small fraction is sufficient to induce strong van der Waals interactions between the sensor and the bath, creating the steep energy penalty $J$. Crucially, if the sensor is in the ground state $|g\rangle$, the laser is far off-resonance, the Rydberg mixing is negligible, and the bath remains non-interacting (shallow $\epsilon \approx 0$). The setup is illustrated in fig. \ref{fig:expsetup}. 

We remark that this setup is slightly different from that described in \eqref{eq:TC}, although the entropic protection mechanism is the same. In \eqref{eq:TC}, for each stabilizer, there is a bath mode carrying $M$ equally
spaced, non-degenerate levels, with a bandwidth of $(M-1)\epsilon$. Here, one stabilizer is associated with $m$ bath atoms, so the weights factorize:
$
w_+=\prod_{a=1}^{m}\big(1+e^{-\beta\epsilon_a}\big),$ and
$w_-=\prod_{a=1}^{m}\big(1+e^{-\beta J_a}\big).
$
With $\beta\epsilon\ll1$ one has $\ln(1+e^{-\beta\epsilon})=\ln2-\beta\epsilon/2
+O(\beta^2\epsilon^2)$, while $\beta J\gg1$ gives $w_-\approx1$; for uniform
$\epsilon_a,J_a$ this yields $\lambda\simeq M e^{-m\beta\epsilon/2}$, so that
saturation requires only $\tfrac{1}{2}m\beta\epsilon\ll1\ll\beta J$ ---
logarithmic in $M$ rather than linear.

The defect penalty $J$ is set by the strength of the Rydberg-dressed interaction between the sensor and bath atoms. For example, the spin-flip blockade in \cite{Jau_2015} was able to realize a defect penalty of $\sim 1$MHz in units of Planck constant. The residual conditional coupling from far-detuned Rydberg dressing is strongly suppressed, yielding $\epsilon/h \sim 10$ kHz for typical parameters.  
The temperature entering $\beta$ here is that of the internal Cs states, which
can be randomized by an applied drive. Appendix \ref{app:Lindblad}
gives the corresponding Lindbladian, whose steady state has the conditional product form of
$w_{\pm}$ and obeys detailed balance. The plateau requires the drive spectrum to cover the residual shifts $\epsilon_a$ but not the blockade shifts $J_a$. The resulting effective bath temperature is set by the bandwidth of that
drive, $k_BT/h\sim200$~kHz at the operating point identified in Appendix~\ref{app:Lindblad},
while the motional temperature and control-field noise enter only through the
bare rate $\Gamma_0$.

At these quoted scales,  $\tfrac12m\beta\epsilon=0.2$ and
$\beta J=5$, such that the saturation requirement $\frac{1}{2}m\beta \epsilon \ll 1\ll \beta J$ is accessible. We remark that while our low temperature requirement to reach the desired code sector is similar to that of other existing state-preparation methods, this alone does not protect against coherent control errors and noise-induced defect diffusion. The entropic stabilization mechanism suppresses logical error processes after preparation, 
and can be understood as a hardware pre-filter, 
lowering the physical error rate $p$ entering a conventional decoder. For a surface-code patch of distance $d$, the number of independent faults
required to produce a logical error is $\lfloor(d+1)/2\rfloor$, so below
threshold $\Gamma_L\sim(p/p_{\rm th})^{\lfloor(d+1)/2\rfloor}$ with
$p_{\rm th}\simeq1\%$ \cite{Dennis_2002,PhysRevA.86.032324}. Fixing a target $\Gamma_L$ set by the intended
computation and solving for the distance required to reach it, with
$\lfloor(d+1)/2\rfloor\to d/2$ and $p\to p/M^{2}$,
\begin{equation}
d\ \simeq\ \frac{2\ln(1/\Gamma_L)}{\ln\!\big(M^{2}p_{\rm th}/p\big)} .
\end{equation}
Taking $\Gamma_L=10^{-10}$ and $p/p_{\rm th}=0.1$, representative of demonstrated
two-qubit fidelities an order of magnitude below threshold, the required distance
falls from $21$ to $5$ at $m=7$, and the total atom count from $881$ to $217$
even after the reservoir overhead \footnote{The estimate is a scaling one: at such small
distances the asymptotic form of $\Gamma_L$ is only indicative, and the
substitution $p\to p/M^{2}$ identifies the suppression of anyon creation with a
reduction of the effective error rate seen by the decoder, which does not extend
to gate or readout errors during syndrome extraction.}. The major constraint is
the fidelity of the stabilizer-mapping circuit, which imprints the stabilizer
eigenvalue onto the sensor atom whose state controls the dressing shift seen by
the reservoir: if it fails with probability $p_{\rm map}$ per cycle the reservoir
is transiently unconfined, the gating probability saturates at
$\max(M^{-1},p_{\rm map})$, and $M\lesssim p_{\rm map}^{-1}$ --- so that $m=7$
requires a mapping fidelity of $99.2\%$, at the level already demonstrated \cite{PhysRevLett.129.030501,Evered_2023}.

\section{Entropic Protection of Superfluids}
\label{sec:XY_Model}

Having demonstrated the stabilization of discrete topological memory, we now extend the framework to continuous variables by considering the two-dimensional XY model coupled to entropic reservoirs. The physics here is governed by the binding and unbinding of vortex defects, described by the Berezinskii--Kosterlitz--Thouless (BKT) transition.

Consider a square lattice of planar rotors $\theta_i \in [0,2\pi)$. The standard XY interaction is dressed by a coupling to local baths located on the plaquettes $p$. The full Hamiltonian is
\begin{equation}
    H = -J \sum_{\langle ij \rangle} \cos(\theta_i - \theta_j) + \sum_p H_p^{\text{bath}}(\Phi_p),
\label{eq:XY}
\end{equation}
where $\Phi_p = \sum_{\square} (\theta_i - \theta_j)$ denotes the vorticity around plaquette $p$, defined using the usual convention that angular differences are taken modulo $2\pi$. 
The bath term $H_p^{\text{bath}}$ imposes an entropic penalty on nonzero vorticity. 

We assume the saturated entropic plateau regime: when $\Phi_p = 0$, the local bath has volume $M$, while for $\Phi_p \neq 0$ the bath is confined to a single microstate. With the lattice constant set to unity, the free
energy of a single vortex in a box of size $L$ is
\begin{equation}
\begin{split}
\beta F_{\rm v}=
& \underbrace{\pi K\ln L}_{\rm elastic}+\underbrace{\beta E_c}_{\rm core}
-\underbrace{2\ln L}_{\rm position}+\underbrace{\ln M}_{\rm bath} \\
= & (\pi K-2)\ln L+\beta E_c+\ln M.\\
\end{split}
\label{eq:Fv}
\end{equation}
A free vortex is favorable once $\beta F_{\rm v}<0$, so on the disordered side
$\pi K<2$ the condition $\beta F_{\rm v}=0$ locates the scale beyond which free
vortices proliferate, i.e. the correlation length: 
$\xi=\xi_0M^{\nu_{\rm eff}}$ with $\xi_0=e^{\beta E_c\nu_{\rm eff}}$ and
$\nu_{\rm eff}=(2-\pi K)^{-1}$. Note that $\ln M$ enters additively alongside
$\beta E_c$, with no factor of $\ln L$: the reservoir shifts the bare vortex fugacity from $y_0=e^{-\beta E_c}$ to
$y_{\rm eff}=y_0/M$, but leaves the bare stiffness $K=\beta J$ untouched. 
Solving $\beta F_{\rm v}=0$ for temperature instead gives the crossover at finite system size (instead of a true critical point),
\begin{equation}
\frac{T_c(L,M)}{T_c(L,1)}=\frac{2\ln L}{2\ln L-\ln M}
=\Big(1-\frac{\ln M}{2\ln L}\Big)^{-1}.
\label{eq:Tc}
\end{equation}
The enhancement decays only as $1/\ln L$, remaining at the $20\%$ level or above
provided $M\gtrsim L^{1/3}$ --- a mild requirement, $M\gtrsim100$ even at
$L=10^{6}$. The denominator vanishes at $M=L^{2}$; beyond that the bath entropy
exceeds the positional entropy of a vortex, no free vortex is favorable at any
temperature, and the finite-size crossover disappears altogether. Equation~\eqref{eq:Tc} 
therefore describes the crossover across the range $L^{1/3}\lesssim M\lesssim
L^{2}$, the upper end marking not a failure of the mechanism but the point
beyond which coherence is no longer lost at any accessible temperature. 
We note that Eq.~\eqref{eq:Fv} uses the bare stiffness and therefore neglects the
dielectric screening of $K$ by bound vortex pairs. The screening is second order in the vortex fugacity, and the reservoir suppresses that fugacity by $M$, so the neglected correction is itself suppressed by $M^{2}$. Eq.~\eqref{eq:Tc}  should nonetheless be read as an estimate of the finite-size coherence scale rather than of the
renormalized transition.

The logarithmic interaction, however, does not affect the kinetic suppression. The system
energy reads
\begin{equation}
E=\frac{J}{2}\!\int\!(\nabla\theta_{\rm sw})^{2}
 -\pi J\!\!\sum_{i\neq j}\!q_iq_j\ln|\bm r_i-\bm r_j|
 +E_c\sum_iq_i^{2},
 \label{eq:XY_E}
\end{equation}
so the Coulomb kernel is fixed by the same stiffness $J$ that governs the
spin-wave sector. Since Eq. \eqref{eq:XY} couples the reservoir to the local vorticity $\Phi_p$ and not to
$\nabla\theta$, the bare stiffness is unchanged and the kernel in
Eq.~\eqref{eq:XY_E} is unmodified. The indirect, $O(y^{2})$ modification of the
renormalized stiffness noted above alters the force between vortices, not
their mobility.
The suppression $M^{-1}$ is the probability that the destination reservoir is found in the state compatible with a core, and it multiplies forward and
backward hops symmetrically, preserving detailed balance. Both the diffusion
constant and the mobility $\mu$  
are therefore reduced by the same factor,
\begin{equation}
D_{\rm eff}=\frac{D_0}{M},\qquad
\mu_{\rm eff}=\frac{\mu_0}{M},\qquad
\mu_{\rm eff}=\frac{D_{\rm eff}}{k_BT} ,
\end{equation}
so that the Einstein relation is intact, and the interacting problem --- pair binding, Coulomb drift, recombination --- is reproduced with the clock rescaled by $t\to t/M$.

We now examine the phase-slip rate $\Gamma_{\mathrm{slip}}$. Let $W$ be the winding number of $\theta$ around a non-contractible loop along $x$, averaged over the $L$ parallel rows of the torus, and define $\Gamma_{\rm slip}$ through $\langle(\Delta W)^2\rangle\simeq\Gamma_{\rm slip}t$ at long times. A $2\pi$ phase slip thus corresponds to $\Delta W=\pm1$.  The winding along a single loop changes by $\pm 1$ every time a vortex crosses it, and a vortex that moves in the normal direction by $\Delta y$ crosses $\Delta y$ of the loops. Hence $\Delta W = L^{-1} \sum_v q_v \Delta y_v$, with $q_v$ the charge of vortex $v$. 
When $T>T_c(L,M)$, free vortices of density $n_f$ are present in equilibrium and diffuse independently, each with $\langle(\Delta y_v)^2\rangle=2D_{\rm eff}t$, while bound pairs contribute only bounded displacements. The cross terms average to zero, giving  $\Gamma_{\text{slip}}\simeq 2n_f D_{\mathrm{eff}}$. Since the dynamics obeys detailed balance, $n_f$ is the equilibrium density. For a dilute, weakly interacting gas, 
each vortex carries the single-vortex weight $y_{\rm eff}$: vortices are created in pairs at a rate $\propto y_{\rm eff}^2$, but they also annihilate in pairs at a rate $\propto n_+n_-$, so that $n_+n_-=y_{\rm eff}^2$. With neutrality, $n_+=n_-$, such that $n_f=n_+ + n_- = 2y_{\mathrm{eff}}+\mathcal{O}(y_{\mathrm{eff}}^2)$. Consequently 
\be
\Gamma_{\mathrm{slip}} \simeq 4 y_{\mathrm{eff}} D_{\mathrm{eff}} = 4y_0 D_0/M^2\sim M^{-2},
\ee
one factor of $M^{-1}$ from the fugacity and one from the mobility. This combined suppression of vortex density and mobility provides a dynamical complement to the thermodynamic enhancement of the correlation length. The logarithmic interaction further strengthens the suppression since $n_f\sim \xi^{-2} \propto M^{-2\nu_{\mathrm{eff}}}$ with $2\nu_{\mathrm{eff}}>1$. And outside this regime, below the crossover $T<T_c(L,M)$, each slip requires a pair to nucleate $(\propto y_{\mathrm{eff}}^2)$ and to separate $(\propto D_{\mathrm{eff}})$, leading to the suppression of $\Gamma_{\mathrm{slip}}\propto M^{-3}$, the analogue of \eqref{eq:TC_lifetime}.

To conclude the section, we comment on the difference between our mechanism and other proposals in the literature. The entropic coupling does not touch the bare stiffness, so that the spin-wave sector is unaffected and there is no true long-range order at any finite $M$ and $T>0$ in the thermodynamic limit. 
The effect is confined to the vortex sector, and the enhancement of Eq. \eqref{eq:Tc} is a finite-size statement, unlike proposals such as \cite{1jkm-kybr} in quasi-2D systems. Additionally, since the $\ln M$ shift is scale-independent, it
amounts to a renormalization of the bare core energy and leaves $\nu_{\rm eff}=(2-\pi K)^{-1}$ unchanged; it is therefore distinct from defect couplings that themselves flow under coarse-graining, as in Refs.  \cite{SciPostPhys.12.4.131,PhysRevLett.128.215701}.

\section{Discussion and outlook}

We have shown that entropy, typically the adversary of quantum order, can be engineered to serve as its protector. By creating a landscape where topological defects are entropically confined, in finite systems one can achieve a polynomial enhancement in toric code memory lifetime ($\tau \sim M^3$) and a significant expansion of superfluid coherence (slip error rate suppressed at least as $\sim M^{-2})$. This mechanism offers a new, resource-efficient paradigm for stabilizing quantum matter in the noisy intermediate-scale era.

Here we comment more on the experimental setup of entropic toric code in section \ref{subsec:experiment}. The primary advantage of this setup is the exponential resource scaling. A modest bath of $m=7$ atoms can already yield a Hilbert-space dimension $M=2^{m}=128$, giving an entropic suppression factor $M^3\simeq 2.1\times 10^6$. Furthermore, the parameters $J$ and $\epsilon$ are optically tunable, allowing dynamic exploration of the phase diagram and the breakdown transition.

A practical limitation of the proposed implementation is the finite coherence time associated with Rydberg dressing. While entropic protection suppresses the generation and transport of logical errors, the protection itself is ultimately limited by photon scattering from the dressing lasers and by the effective lifetime of the weakly mixed Rydberg states, typically in the range of $10-100$ ms 
\cite{PhysRevLett.104.195302}.
This finite lifetime is not unique to our setup but reflects a general constraint of Rydberg-based implementations; indeed, the timescale achievable with Rydberg dressing is often longer than the coherence window in experiments based on resonant Rydberg blockade. Another limitation is that the scheme relies on the fidelity of the stabilizer-mapping circuit that conditions the bath potential on the stabilizer eigenvalue. Imperfections in gate operations or sensor readout effectively introduce leakage channels that reduce the entropic barrier. Encouragingly, recent Rydberg experiments have demonstrated multi-qubit parity and stabilizer-mapping fidelities above the 99\% level (see for example \cite{PhysRevLett.129.030501,Evered_2023}), indicating that this requirement is compatible with current capabilities.

An intriguing direction for future work is the hybridization of passive entropic protection with active quantum error correction. In conventional active schemes, a decoder must identify and correct errors faster than they accumulate, requiring the physical error rate to lie below a threshold 
$p<p_c$. Entropic protection can be viewed as a hardware-level pre-filter that renormalizes the effective rate of dangerous error processes. Operating a standard decoder on top of an entropically protected substrate could therefore enable fault tolerance using significantly noisier physical components.

More broadly, this work also raises the question of whether entropic baths can be used to stabilize conformal critical points at finite temperature. For example in the transverse-field Ising chain, the critical point is governed at zero temperature by the Ising conformal field theory, while finite temperature normally produces a quantum critical fan rather than a true critical state. It would be interesting to explore whether an entropic bath can effectively counteract thermal decoherence and preserve conformal scaling over extended length and time scales. Such scenarios would represent a qualitatively new form of bath-stabilized criticality, which may be related to the investigations in Refs.~\cite{PhysRevLett.125.131603,PhysRevD.102.065014,10.21468/SciPostPhys.14.6.168,Chai_2021,Chaudhuri_2021,PhysRevD.103.096014,Chaudhuri_2021, Agrawal_2021, PhysRevLett.134.041602, yb7d-6tvc, PhysRevD.103.026008, BUCHEL2021115425, BUCHEL2021115605, BUCHEL2024116578, 10.21468/SciPostPhys.12.6.181}.

\begin{acknowledgments}
We thank Yiqiu Han for introducing her work, Itamar Kimchi and Xueda Wen for discussions related to Appendix \ref{app:modified_Ising}, and Google Gemini 3 Pro for inspiring conversations. This research was supported in part by grant NSF PHY-2309135 to the Kavli Institute for Theoretical Physics (KITP). 
\end{acknowledgments}

\appendix
\section{Modified Ising chain}
\label{app:modified_Ising}
\begin{figure}[htbp]
\centering
\includegraphics[scale=0.5]{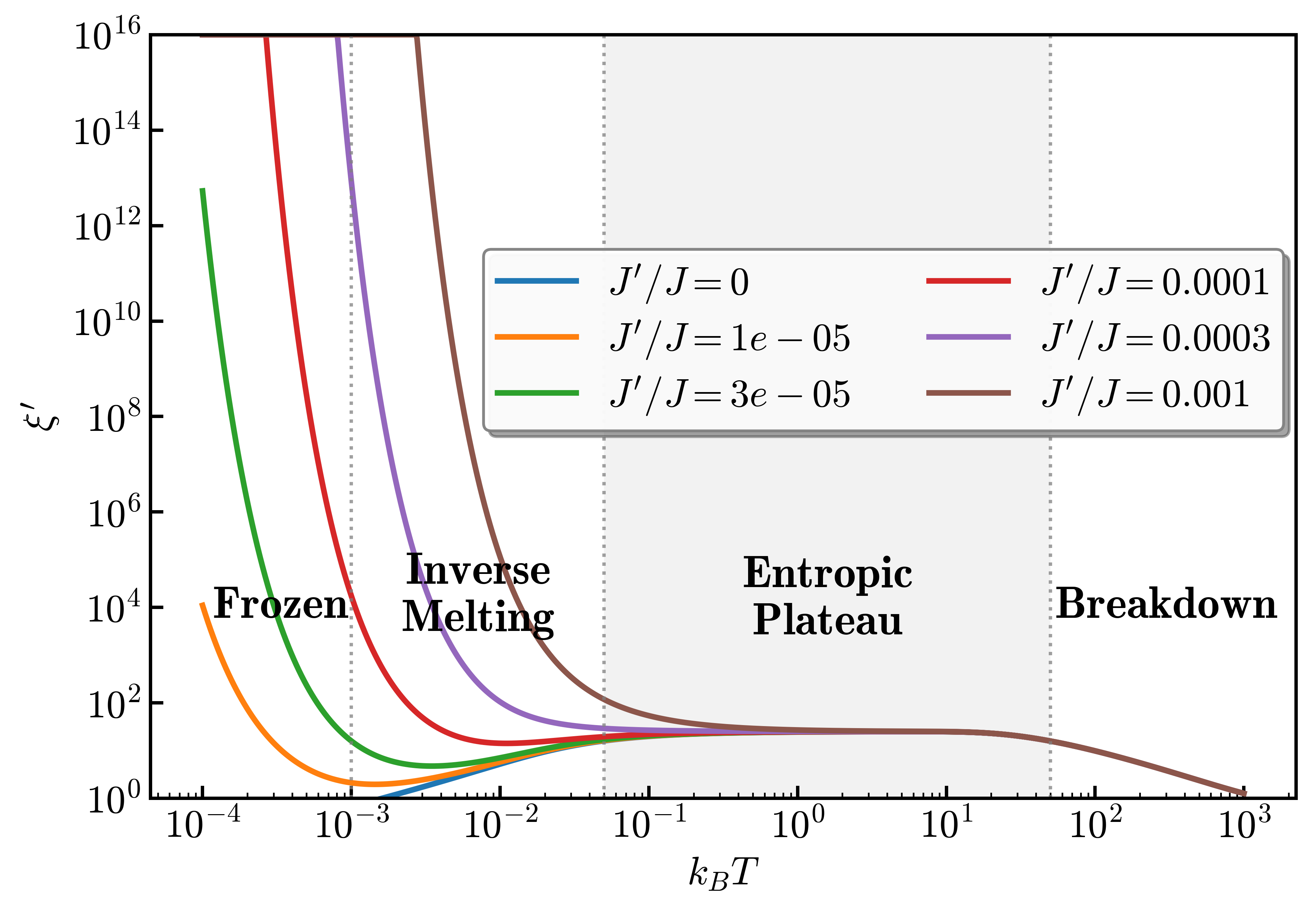}
\caption{Correlation length as a function of thermal energy for different values of $J'$.}
\label{fig:modified_Ising}
\end{figure}

In this Appendix, we connect the entropic stabilization physics \eqref{eq:Ising} to that of the conventional Ising chain by adding
\be
H' = H - J' \sum_i Z_i Z_{i+1}.
\ee
This term does not modify the structure of the transfer-matrix solution, but simply renormalizes the ratio~\eqref{eq:lambda} as
\begin{equation}
\lambda' = \lambda\, e^{2\beta J'} .
\end{equation}
Consequently, the correlation length becomes 
\be
\xi' =\left[ \ln \frac{\lambda' +1}{\lambda' -1}\right]^{-1}.
\ee
In the low temperature limit $\beta\rightarrow \infty$, the additional exponential factor $e^{2\beta J'}$ becomes enormous, such that $\xi'$ diverges as expected in the standard Ising model. At higher temperatures, as long as $\beta J'$ is small compared with other scales in the system, the entropic plateau region is still intact. In fig. \ref{fig:modified_Ising}, we plot the correlation length plateau as a function of $J'$ and thermal energy.

\section{Microscopic Derivation of Entropic Lindbladian}
\label{app:Lindblad}

In this appendix, we derive the effective Lindblad master equation governing the memory dynamics. We start from the microscopic Hamiltonian of the toric code coupled to local entropic reservoirs and perform an adiabatic elimination of the bath degrees of freedom under the assumption of weak system--bath coupling. We model physical error processes as a weak coupling between the system qubits and an external, unbounded environment, mediated by single-qubit operators. The interaction Hamiltonian is
\begin{equation}
    V = \sum_l \left( X_l \otimes \mathcal{E}_l^x + Z_l \otimes \mathcal{E}_l^z \right),
\end{equation}
where $\mathcal{E}_l^{x,z}$ are operators acting on the external environment. We treat this perturbatively using Fermi's Golden Rule to derive transition rates between stabilizer configurations. Crucially, the interaction acts only on the system qubits and leaves the local entropic bath states unchanged.

Consider a transition from a stabilizer configuration $s$ to $s'$ induced by a bit-flip $X_l$. The corresponding transition rate is
\begin{align}
    \Gamma_{s \to s'} = & \sum_{\{n\}_{s'},\{n\}_s}
    \mathcal{P}(\{n\} | s) \Gamma_{\mathrm{env}}(\omega)\nonumber \\
    & \times \left|
    \langle \{n\}_{s'} | \langle s' | X_l | s \rangle | \{n\}_s \rangle
    \right|^2 ,
\label{eq:transition_rate}
\end{align}
where $\mathcal{P}(\{n\} | s)$ denotes the stationary probability of finding the bath configuration $\{n\}$ conditioned on the stabilizer configuration $s$, and $\Gamma_{\mathrm{env}}(\omega)=\hbar^{-2}\int dt e^{-\mathrm{i}\omega t}\langle \mathcal{E}_l^x (t) \mathcal{E}_l^x (0)\rangle$ is the rate at which the environment delivers energy $\hbar \omega = E_{s'}(\{n\})-E_s (\{n\})$ to the system. $\mathcal{E}_l^z$ is taken to have the same noise spectrum. 
Since $X_l$ acts trivially on the bath degrees of freedom, the matrix element is nonzero only if the bath configuration is conserved, $\{n\}_{s'}=\{n\}_s$. Each plaquette whose stabilizer flips from $+1$ to $-1$ thus changes the energy by $(J-\epsilon)n_p$, or by $\sum_a (J_a - \epsilon_a) n_{p,a}$ for the factorized bath described in Sec. \ref{subsec:experiment}. We require the environment to be white across the resonant line:  $\Gamma_{\mathrm{env}}\simeq \Gamma_0$ for $|\omega|$ up to the rate $\sim m\gamma_{\rm bath}$ at which an empty reservoir is repopulated (see below), but unable to supply the defect penalty, i.e. $\sum_a \Gamma_{\mathrm{env}} \big( (J_a-\epsilon_a)/\hbar \big)\ll \Gamma_0$. For a thermal environment at inverse temperature $\beta$, the above requirement reduces to $\beta J\gg1$; flips that release the penalty remain allowed, but they start only from configurations of weight $w_--1\ll1$ and are neglected. The condition is also met naturally by classical control noise, such as fluctuations of magnetic fields and trap-laser intensities, whose spectra are typically concentrated well below $(J-\epsilon)/h\approx1$~MHz. 
For noise satisfying the condition, only configurations with $n_p=0$ on every plaquette whose stabilizer flips contribute, each at the rate $\Gamma_0$.
 Consequently, the transition rate is $\Gamma_0$ times the probability that the bath configuration compatible with $s$ costs no energy in $s'$.

\paragraph{Pair creation.}
Both plaquettes $p_1$ and $p_2$ initially lie in the target sector ($B_p=+1$) and are flipped into the defect sector ($B_p=-1$). In the entropic plateau regime, the initial bath configurations are uniformly distributed, so that $\mathcal{P}(n_{p_1},n_{p_2}) \approx M^{-2}$. In the final configuration, the steep potential enforces $n_{p_1}=n_{p_2}=0$; any other configuration incurs an energetic penalty $J$. The transition is therefore resonant only if the initial bath configuration already satisfies this constraint, yielding a pair-creation rate
\begin{equation}
    \Gamma_{\mathrm{cr}} \approx \Gamma_0 M^{-2}.
\label{eq:Gamma_cr}
\end{equation}

\paragraph{Anyon diffusion.}
This process corresponds to one plaquette $p_1$ relaxing ($-\!\to\!+$) while the neighboring plaquette $p_2$ is excited ($+\!\to\!-$). The initial bath probability is $\mathcal{P}(0,n_{p_2}) \approx M^{-1}$. After the hop, the bath at $p_1$ experiences a flat potential, while the bath at $p_2$ must satisfy $n_{p_2}=0$. The resulting diffusion rate is
\begin{equation}
    \Gamma_{\mathrm{dif}} \approx \Gamma_0 M^{-1}.
\end{equation}

\paragraph{Pair annihilation.}
In this process both plaquettes relax back to the vacuum sector. Initially, the baths are frozen at $(n_{p_1},n_{p_2})=(0,0)$ with probability $\mathcal{P}(0,0)\approx 1$. In the final configuration, the potential is flat and all bath states are allowed, so the annihilation rate is unsuppressed:
\begin{equation}
    \Gamma_{\mathrm{ann}} \approx \Gamma_0.
\label{eq:Gamma_ann}
\end{equation}

Although physical noise acts uniformly on all links, the entropic reservoirs selectively filter these events depending on the instantaneous stabilizer configuration. To capture this state-dependent dynamics, we decompose the bit-flip operator $X_l$ acting on the link shared by plaquettes $p_1$ and $p_2$ into transition operators between specific stabilizer subspaces. We define projectors onto the vacuum (target) and defect (error) subspaces of a plaquette $p$ as
\begin{equation}
    \Pi_p^{+} = \frac{1+B_p}{2}, \qquad \Pi_p^{-} = \frac{1-B_p}{2}.
\end{equation}
The operator $X_l$ can then be decomposed into three distinct jump operators corresponding to the three processes discussed above:
\begin{equation}
\begin{split}
& L_{\mathrm{cr},X_l} = X_l \Pi_{p_1}^{+} \Pi_{p_2}^{+}= \Pi_{p_1}^{-} \Pi_{p_2}^{-} X_l, \\
& L_{\mathrm{dif},X_l}^{\to 1} = X_l \Pi_{p_1}^{-} \Pi_{p_2}^{+}, \quad L_{\mathrm{dif},X_l}^{\to 2} = X_l \Pi_{p_1}^{+} \Pi_{p_2}^{-},\\
& L_{\mathrm{ann},X_l} = X_l \Pi_{p_1}^{-} \Pi_{p_2}^{-}= \Pi_{p_1}^{+} \Pi_{p_2}^{+} X_l .
\end{split}
\end{equation}
The total effective Lindbladian is obtained by summing over all links $l$ (and analogously including $Z_l$ errors on the vertices):
\begin{equation}
\begin{split}
\dot{\rho} = \sum_l \big(
& \Gamma_{\mathrm{cr}} \mathcal{D}[L_{\mathrm{cr},X_l}]
+ \Gamma_{\mathrm{dif}} \big( \mathcal{D}[L_{\mathrm{dif},X_l}^{\to 1}]
+ \mathcal{D}[L_{\mathrm{dif},X_l}^{\to 2}] \big) \\
& + \Gamma_{\mathrm{ann}} \mathcal{D}[L_{\mathrm{ann},X_l}]
+ (X_l \to Z_l) \big) \rho ,
\end{split}
\end{equation}
where $\mathcal{D}[L]\rho = L\rho L^\dagger - \frac{1}{2}\{L^\dagger L,\rho\}$. Note that when replacing $X_l\rightarrow Z_l$, the projectors are correspondingly replaced by $\Pi_v^{\pm}=(1\pm A_v)/2$. 
Substituting the rates derived above yields the final master equation,
\begin{equation}
\begin{split}
\dot{\rho} = \Gamma_0 \sum_l \big(
& \frac{1}{M^2} \mathcal{D}[L_{\mathrm{cr},X_l}]
+ \frac{1}{M} \big( \mathcal{D}[L_{\mathrm{dif},X_l}^{\to 1}]
+ \mathcal{D}[L_{\mathrm{dif},X_l}^{\to 2}] \big) \\
& + \mathcal{D}[L_{\mathrm{ann},X_l}]
+ (X_l \to Z_l) \big) \rho .
\end{split}
\end{equation}
This form makes explicit the hierarchy of timescales induced by the entropic reservoirs.

We remark that the stationary distribution of bath $\mathcal{P}(\{n\} |s)$  in \eqref{eq:transition_rate}, conditioned on the system stabilizers, is engineered by a drive on the bath atoms. We use $n_{xa}$ to denote the occupation of bath atom $a$ of a plaquette or vertex $x$, with the Hamiltonian $H_{\mathrm{bath}}=\sum_{x,a} (\epsilon_a \Pi_x^+ + J_a \Pi_x^-)n_{xa}$. The drive is governed by $H_d(t)=\frac{\hbar}{2}\sum_{x,a} [\Omega (t) \sigma_{xa}^+ + \text{h.c.}]$, where $\Omega(t)$ is the complex Rabi frequency of the drive in the rotating frame, from which $\epsilon_a$ and $J_a$ are measured. We take it to be a classical stationary random process with zero mean and random phase, $\langle\Omega(t)\rangle=\langle\Omega(t)\Omega(t')\rangle=0$, so that up to second order it is characterized entirely by its power spectrum $S_\Omega(\omega)=\int d\tau\,e^{i\omega\tau}\langle\Omega(\tau)\Omega^*(0)\rangle$, with $\int S_\Omega(\omega)\,d\omega/2\pi=\langle|\Omega|^2\rangle$. When the rms Rabi frequency $\langle|\Omega|^2\rangle^{1/2}$ is small compared with the inverse correlation time of $\Omega(t)$, i.e.\ with the half-width $B$ of $S_\Omega$, the drive can be eliminated to second order in $\Omega$. 

We first consider a single bath atom $a$ in a fixed sector $s$, since the bath dynamics conserve the stabilizers ($\Pi^\pm_x$ commute with $H_{\rm bath}$ and $H_d$). With $n=\sigma^+\sigma^-$, the atom's Hamiltonian is $\hbar \Delta_s n$, with $\Delta_+=\epsilon_a/\hbar$ and $\Delta_-=J_a/\hbar$. In the interaction picture with respect to $\hbar\Delta_s n$, $\sigma^\pm\to e^{\pm \mathrm{i}\Delta_s t}\sigma^\pm$, and the drive Hamiltonian becomes
\begin{equation}
\tilde H_{d,s}(t)=\frac{\hbar}{2}\big[\Omega(t)e^{\mathrm{i}\Delta_s t}\sigma^++\Omega^*(t)e^{-\mathrm{i}\Delta_s t}\sigma^-\big].
\end{equation}
Integrating $\dot{\tilde\rho}=-\frac{\mathrm{i}}{\hbar}[\tilde H_{d,s}(t),\tilde\rho]$ from the initial time $t_0$ and inserting the result back into the right-hand side gives
\begin{equation}
\begin{split}
\dot{\tilde\rho}(t)=&-\frac{\mathrm{i}}{\hbar}\big[\tilde H_{d,s}(t),\tilde\rho(t_0)\big]\\
&-\frac{1}{\hbar^2}\int_{t_0}^{t}dt'\,\big[\tilde H_{d,s}(t),[\tilde H_{d,s}(t'),\tilde\rho(t')]\big].
\end{split}
\end{equation}
Upon averaging over the noise, the first term vanishes because $\langle\Omega\rangle=0$. For the second term, the correlations of $\tilde H_{d,s}$ decay within a time of order $B^{-1}$, during which the drive rotates the atom by an angle of order $\langle|\Omega|^2\rangle^{1/2}/B\ll1$,  i.e.\ transfers a probability of order $\langle|\Omega|^2\rangle/B^2$. 
Since the phases of $\Omega(t)$ in successive intervals are independent, these probabilities add, and with $\sim B$ intervals per unit time the averaged state $\rho\equiv\langle\tilde\rho\rangle$ changes at rates of order $\langle|\Omega|^2\rangle/B$.  To leading order in $\langle|\Omega|^2\rangle/B^2$, one may thus average $\tilde H_{d,s}(t)\tilde H_{d,s}(t')$ independently of $\tilde\rho(t')$. The latter can then be replaced by $\rho(t)$ upon averaging, and, for $t-t_0\gg B^{-1}$, one can change variable $\tau=t-t'$ and extend the integral to infinity:
\begin{equation}
\dot\rho=-\frac{1}{\hbar^2}\int_0^\infty d\tau\,\big\langle\big[\tilde H_{d,s}(t),[\tilde H_{d,s}(t-\tau),\rho]\big]\big\rangle,
\label{eq:B12}
\end{equation}
where the remaining average acts on the drive only. Define the correlator $C(\tau)\equiv\langle\Omega(\tau)\Omega^*(0)\rangle$ which satisfies $C(-\tau)=C^*(\tau)$. Using stationarity, the averaged double commutator in \eqref{eq:B12} then evaluates to 
\begin{equation}
\frac{\hbar^2}{4}\Big(C(\tau)e^{\mathrm{i}\Delta_s\tau}[\sigma^+,[\sigma^-,\rho]]
+C^*(\tau)e^{-\mathrm{i}\Delta_s\tau}[\sigma^-,[\sigma^+,\rho]]\Big).
\end{equation}
The integral over $\tau$ thus produces $G(\Delta_s)=\int_0^{\infty} d\tau C(\tau) e^{\mathrm{i}\Delta_s \tau}$ and its complex conjugate. Using again $C(-\tau)=C^*(\tau)$, the sum $G(\Delta)+G^*(\Delta)=S_\Omega(\Delta)$. The difference reads $G(\Delta)-G^*(\Delta)=2i\Lambda(\Delta)$, with the principal-value integral $\Lambda(\Delta)=\int\frac{d\omega}{2\pi}\frac{S_\Omega(\omega)}{\Delta-\omega}$. Further expanding the double commutators leads to 
\begin{equation}
\begin{split}
[\sigma^+,[\sigma^-,\rho]]+[\sigma^-,[\sigma^+,\rho]]&=-2\big(\mathcal{D}[\sigma^+]+\mathcal{D}[\sigma^-]\big)\rho,\\
[\sigma^+,[\sigma^-,\rho]]-[\sigma^-,[\sigma^+,\rho]]&=[\sigma^z,\rho],
\end{split}
\end{equation}
Collecting all terms, we arrive at
\begin{equation}
\begin{split}
\dot\rho&=-\frac14\Big(G(\Delta_s)[\sigma^+,[\sigma^-,\rho]]+G^*(\Delta_s)[\sigma^-,[\sigma^+,\rho]]\Big)\\
&=\frac{S_\Omega(\Delta_s)}{4}\big(\mathcal{D}[\sigma^+]+\mathcal{D}[\sigma^-]\big)\rho-\frac{\mathrm{i}}{4}\Lambda(\Delta_s)[\sigma^z,\rho].
\end{split}
\label{eq:onebath}
\end{equation}
The last term in Eq.~\eqref{eq:onebath} is an AC Stark shift of the transition energy by $\hbar \Lambda(\Delta_s)/2$ and can be absorbed into $\epsilon_a$ and $J_a$.

For the full bath, the only term of $H_{\rm bath}$ that does not commute with $\sigma^+_{xa}$ is $(\epsilon_a\Pi^+_x+J_a\Pi^-_x)\,n_{xa}$, whose frequency depends on the sector. In the interaction picture, therefore,
\begin{equation}
\sigma^+_{xa}\to e^{\mathrm{i}(\epsilon_a\Pi^+_x+J_a\Pi^-_x)t/\hbar}\,\sigma^+_{xa}=\sum_s\Pi^s_x\sigma^+_{xa}\,e^{\mathrm{i}\Delta^{(a)}_st}.
\end{equation}
The second-order generator then reproduces Eq.~\eqref{eq:onebath} within each sector. The cross terms between two sectors of the same atom, which oscillate at $(J_a-\epsilon_a)/\hbar$, large compared with all rates, drop out in the secular approximation. The cross terms between different atoms will be discussed in the next paragraph. Further adding a weak relaxation of every bath atom into $n_{xa}=0$ at a rate $\gamma_\downarrow$ independent of the sector, we obtain
\begin{equation}
\begin{split}
\mathcal{L}_{\rm bath}\rho=\sum_{x,a}\Big[&\sum_{s=\pm}\gamma^{(a)}_s\big(\mathcal{D}[\Pi^s_x\sigma^+_{xa}]+\mathcal{D}[\Pi^s_x\sigma^-_{xa}]\big)\\
&+\gamma_\downarrow\mathcal{D}[\sigma^-_{xa}]\Big]\rho,
\end{split}
\label{eq:Lbath}
\end{equation}
with $\gamma^{(a)}_s=S_\Omega(\Delta^{(a)}_s)/4$. For a power spectrum flat across the small detunings $\epsilon_a/\hbar$ of the satisfied sector $s=+$, which for a band of half-width $B$ requires  $\epsilon_a\lesssim\hbar B$, all atoms in that sector flip at the same rate $\gamma^{(a)}_+\simeq\gamma_{\rm bath}\equiv S_\Omega(0)/4$.

We assume that different bath atoms flip independently, as the jump operators of Eq.~\eqref{eq:Lbath} presuppose. This is automatic for independent noise on each atom; for a common field $\Omega(t)$, it requires the transition frequencies of the atoms within a cluster to differ by more than $\gamma_{\rm bath}$, e.g.\ through their unequal distances to the sensor or site-resolved AC Stark shifts inside the drive band, so that the cross terms between them oscillate and drop out in the secular approximation. Otherwise the field flips degenerate atoms collectively, confining an emptied reservoir to its $(m+1)$-dimensional permutation-symmetric subspace, until $\gamma_\downarrow$ breaks the symmetry. In this case, the bath re-equilibrates at a rate of order $\gamma_\downarrow$ rather than $\gamma_{\rm bath}$. By comparison, degeneracies between different clusters are tolerable, since they only correlate two reservoirs released at the same moment, and only until $\gamma_\downarrow$ acts.

Within a sector, denote the populations of each atom as  $p_n$, with $\dot p_1=\gamma^{(a)}_sp_0-(\gamma^{(a)}_s+\gamma_\downarrow)p_1$. Defining the stationary occupation ratio  $p_1/p_0=r^{(a)}_s$, it is then straightforward to show $p_n=(r^{(a)}_s)^n/(1+r^{(a)}_s)$ for $n\in\{0,1\}$. Since Eq.~\eqref{eq:Lbath} is a sum of single-atom generators, which commute, its stationary state in sector $s$ is the product of these distributions, 
\begin{equation}
\mathcal{P}(\{n\}|s)=\prod_a\frac{\big(r^{(a)}_s\big)^{n_a}}{1+r^{(a)}_s},\qquad r^{(a)}_s=\frac{\gamma^{(a)}_s}{\gamma^{(a)}_s+\gamma_\downarrow}.
\label{eq:Pbath}
\end{equation}
This is the conditional Gibbs form of Sec.~\ref{subsec:experiment} with $e^{-\beta\epsilon_a}\to r^{(a)}_+$ and $e^{-\beta J_a}\to r^{(a)}_-$.

To find the joint stationary distribution of the system and bath, $\pi(s,\{n\})=P(s)\,\mathcal{P}(\{n\}|s)$, we impose detailed balance. Bath flips leave $s$ unchanged and are balanced by Eq.~\eqref{eq:Pbath}. Under the conditions stated below Eq.~\eqref{eq:transition_rate}, a system flip $s\to s'$ requires $n_{xa}=0$ for all atoms in the set $F$ of sites whose stabilizers change. Such flips happen at the rate $\Gamma_0$ in both directions, so that detailed balance requires $\pi(s,\{n\})=\pi(s',\{n\})$. By Eq.~\eqref{eq:Pbath}, each empty site $x\in F$ contributes a factor $1/w_{s_x}$ to $\mathcal{P}(\{n\}|s)$, where $w_s=\prod_a(1+r^{(a)}_s)$, while all other sites contribute equally to both $\pi (s,\{n\})$ and $\pi (s',\{n\})$. Hence $P(s)/\prod_{x\in F}w_{s_x}=P(s')/\prod_{x\in F}w_{s'_x}$, which is solved by $P(s)\propto\prod_xw_{s_x}$, the entropic Gibbs measure. With all transitions balanced, the stationary distribution of the joint dynamics is
\begin{equation}
\pi(s,\{n\})\propto\prod_{x,a}\big(r^{(a)}_{s_x}\big)^{n_{xa}}.
\end{equation}
In general, equations ~\eqref{eq:Gamma_cr}--\eqref{eq:Gamma_ann} thus read $\Gamma_{\rm cr}=\Gamma_0/w_+^2$, $\Gamma_{\rm dif}=\Gamma_0/(w_+w_-)$ and $\Gamma_{\rm ann}=\Gamma_0/w_-^2$.
 The saturated plateau corresponds to $w_+\simeq M$ and $w_-\simeq1$, 
which is equivalent to requesting 
\be
\sum_a\gamma^{(a)}_-\ll\gamma_\downarrow\ll2\gamma_{\rm bath}/m.
\ee
In other words, the drive must dominate the relaxation in the $s=+$ satisfied sector, keeping each atom close to $p_0=p_1=\frac12$, while the relaxation must dominate the drive in the violated sector $s=-$, holding each atom in $n_{xa}=0$. In terms of the power spectrum, the first inequality reads $\sum_aS_\Omega(J_a/\hbar)/S_\Omega(0)\ll\gamma_\downarrow/\gamma_{\rm bath}$.

Finally we address the separation of timescales that Eq.~\eqref{eq:transition_rate}
presupposes. Conditioning the bath distribution $\mathcal{P}(\{n\} | s)$ on the
instantaneous stabilizer configuration assumes the reservoir re-equilibrates
between successive events at a site, which, by the single-atom rate equation above, happens at the rate $2\gamma_{\rm bath}+\gamma_\downarrow\simeq2\gamma_{\rm bath}$ in the satisfied sector.
The binding
process is the immediate reverse hop: after a defect leaves plaquette $p$, that
reservoir sits in $\hat n=0$ within a now-flat potential, and until it spreads,
the return hop is ungated, proceeding at $\Gamma_0$ rather than $\Gamma_0/M$.
The weight on $\hat n=0$ decays as $M^{-1}(1+e^{-2\gamma_{\rm bath}t})^m$, so requiring it to reach $O(M^{-1})$ within the ungated time $\Gamma_0^{-1}$ gives $2\gamma_{\rm bath}\gtrsim\Gamma_0\ln m$, which we strengthen to the conservative $\gamma_{\rm bath}\gg\Gamma_0\ln M$. 
Furthermore, the drive must cover the spread of residual shifts $\epsilon_a$, which
is itself $O(\epsilon)$, so
$\hbar B\gtrsim\epsilon$; and it must not reach the blockade shifts $J_a$ of the confined sector, so $\hbar B\ll J$. Collecting,
\begin{equation}
\hbar\Gamma_0\ln M\ll\hbar\gamma_{\rm bath}\ll\hbar B,\qquad\epsilon\lesssim\hbar B\ll J.
\label{eq:window}
\end{equation}
With $\Gamma_0\sim10$--$10^{2}\,\mathrm{s^{-1}}$ and $M=128$ one has
$\Gamma_0\ln M\approx 5\times10^{2}\,\mathrm{s^{-1}}$. Since $\epsilon/h\sim10$~kHz and $J/h=1$~MHz, the window is therefore
$10~{\rm kHz}\lesssim B/2\pi\ll1~{\rm MHz}$, 
spanning two orders of magnitude. 
Global microwave driving of the cesium hyperfine states at $\Omega/2\pi=24.6$~kHz in a tweezer array~\cite{Manetsch_2025}, with this rms Rabi frequency spread over a flat band of half-width $B/2\pi\sim200$~kHz, gives $\gamma_{\rm bath}=\pi\langle|\Omega|^2\rangle/4B\approx1.5\times10^4\,{\rm s}^{-1}$, comfortably between $\Gamma_0\ln M$ and $B$, while the much stronger Raman driving of Ref.~\cite{PhysRevA.105.032618}  is not required. Identifying $k_BT\simeq\hbar B$ for the bath sector, the value $k_BT/h\sim200$~kHz used in Sec.~\ref{subsec:experiment} sits near the center of the window.

\setlength{\bibsep}{-1.2pt} 
\bibliography{ref.bib}

\end{document}